\documentstyle[preprint,aps]{revtex}

\begin{document}

\title{Critical dimensions for random walks on random-walk chains}
       
\author{Savely Rabinovich, H. Eduardo Roman, Shlomo Havlin,
        and Armin Bunde}

\address{Minerva Center and Department of Physics, Bar-Ilan University,
         52900 Ramat Gan, Israel\\ and\\
         Institut f\"ur Theoretische Physik, Universit\"at Gie{\ss}en,
         Heinrich-Buff-Ring 16, D-35392 Gie{\ss}en, Germany}

\date{received \today}

\maketitle

\begin{abstract}
The probability distribution of random walks on linear
structures generated by random walks in $d$-dimensional space,
$P_d(r,t)$, is analytically studied for the case
$\xi\equiv r/t^{1/4}\ll1$. It is shown to obey the scaling
form $P_d(r,t)=\rho(r) t^{-1/2} \xi^{-2} f_d(\xi)$, where
$\rho(r)\sim r^{2-d}$ is the density of the chain. Expanding
$f_d(\xi)$ in powers of $\xi$, we find that there exists an
infinite hierarchy of critical dimensions, $d_c=2,6,10,\ldots$, 
each one characterized by a logarithmic correction in $f_d(\xi)$.
Namely, for $d=2$, $f_2(\xi)\simeq a_2\xi^2\ln\xi+b_2\xi^2$;
for $3\le d\le 5$, $f_d(\xi)\simeq a_d\xi^2+b_d\xi^d$; for $d=6$,
$f_6(\xi)\simeq a_6\xi^2+b_6\xi^6\ln\xi$; for $7\le d\le 9$,
$f_d(\xi)\simeq a_d\xi^2+b_d\xi^6+c_d\xi^d$; for $d=10$,
$f_{10}(\xi)\simeq a_{10}\xi^2+b_{10}\xi^6+c_{10}\xi^{10}\ln\xi$,
{\it etc.\/}  In particular, for $d=2$, this implies that the 
temporal dependence of the probability density of being close to
the origin $Q_2(r,t)\equiv P_2(r,t)/\rho(r)\simeq t^{-1/2}\ln t$.
\end{abstract}

\bigskip
Pacs: 5.40.+j, 05.60.+w, 66.30.-h

\newpage
\widetext

\section{Introduction}

Random fractals represent useful models for a variety of disordered
systems found in Nature.  In addition to
their structural properties, fractals have attracted much attention
in recent years because of their interesting transport properties
\cite{Orbach,Havlin,Kehr,Fractal1}.

Of particular interest is the question of how the probability density
of random walks, $P_d(r,t)$, is changed on fractal structures with
respect to its Gaussian form valid on regular $d-$dimensional systems, 
$P_d(r,t)\sim t^{-d/2}\exp(-{\rm const}\times\eta^2)$, where
$\eta=r/t^{1/2}$.  The form of $P_d(r,t)$ on fractals has been
extensively studied in the asymptotic limit $\xi=r/t^{1/d_w}\gg1$
\cite{Havlin,Guyer,Procaccia,BHR,Blumen,Roman92,Aharony92,Roman95,Yuste95},
where $d_w$ is the anomalous diffusion exponent characterizing the
time behavior of the random walks, $\langle r^2(t)\rangle\sim t^{2/d_w}$.
As a result of these investigations, it is now generally accepted that
$P_d(r,t)$ displays a stretched Gaussian form 
\begin{equation}\label{Prtrw1}
P_d(r,t)\sim\rho(r)t^{-d_s/2}\exp(-{\rm const}\times\xi^u),
\qquad \xi\gg 1,
\end{equation}
where $\rho(r)\sim r^{d_f-d}$ is the density of the fractal structure,
$d_f$ is the fractal dimension, $d_s=2 d_f/d_w$ is the spectral
dimension \cite{Orbach}, $u=d_w/(d_w-1)$, and is normalized according
to $\int dr\,r^{d-1}P_d(r,t)=1$.  However, much less is known about
the behavior of $P_d(r,t)$ in the opposite limit when $\xi$ approaches
zero. 

In this paper we concentrate on diffusion in linear random fractal
structures generated by random walks (random-walk chains, RWC) in
$d$-dimensional systems, where $P_d(r,t)$ can be obtained exactly. 
Recently, using numerical simulations, it has been suggested that
for such linear fractals \cite{DRB95},
\begin{equation}\label{Prtrw}
Q_d(r,t)/Q_d(0,t)\sim (1-{\rm const}\times\xi^{d-2}),\qquad \xi\to 0,
\end{equation}
for all dimensions $d$, where $Q_d(r,t)=\rho(r)^{-1} P_d(r,t)$ is
normalized on the fractal chain, {\it i.e.,\/}
$\int dr\,r^{d_f-1} Q_d(r,t)=1$, with $d_f=2$, $d_w=4$ for RWC,
$\xi=r/t^{1/4}$ and $Q_d (0,t)$ is the probability density to
return to the origin.

In the following, we derive an exact expansion for $P_d(r,t)$ in
the limit of $\xi\to0$.  Surprisingly, $P_d(r,t)$ displays an
extremely rich behavior as a function of both $\xi$ and
dimensionality $d$. We show, among other results, that
Eq.~(\ref{Prtrw}) can only be valid for $3\le d\le 5$, and
\begin{equation}\label{Prtrw2}
P_d(r,t)\sim\rho(r)t^{-1/2}(1-{\rm const}\times\xi^4), 
\quad {\rm for} \quad d\ge 7.
\end{equation}
Moreover, we find that the small-$\xi$ 
expansion of $P_d(r,t)$ is characterized by
a hierarchy of cri\-tical dimensions, $d_c=2,6,10,14,\ldots$, where
logarithmic corrections of the form $\xi^{d_c-2}\log(1/\xi)$ occur.
In particular, for $d=2$ we obtain 
$P_2(r,t)\simeq 2\rho(r)t^{-1/2}\ln(t^{1/4}/r)$.

\section{Random walks on random-walk chains}

We consider linear structures generated by random walks 
in $d$-dimensional systems.  Such structures are fractals with 
fractal dimension $d_f=2$, independently of $d$.  To study diffusion
of particles along such linear chains, we assume that the
diffusing particles (random walkers) can move only along the
structure (path) which has been created sequentially by the
generating walks.  Thus, although the structure can intersect
itself in space, the walkers see just a linear path. We denote
such paths as random-walk chains (RWC).

Along the linear path, the probability density of random
walkers, at chemical distance $\ell$ along the RWC from their
starting point after time $t$, $p(\ell,t)$, subject to the
initial condition $p(\ell,0)=\delta(\ell)$, approaches the
well-known Gaussian distribution
\begin{equation}\label{Plt}
p(\ell,t)={2\over(2\pi t)^{1/2}}\exp\left(-{\ell^2\over 2t}\right),
\end{equation}
normalized according to $\int_0^\infty d\ell\,p(\ell,t)=1$.  Thus,
diffusion along the chain ({\it i.e.,\/} $\ell$-space) is normal
and $\langle\ell^2\rangle=t$. On the contrary, in Euclidean
$r$-space diffusion is anomalous with $d_w=2 d_f=4$ (see,
{\it e.g.,\/} \cite{Havlin}).

To obtain the behavior of the probability density in $r$-space,
averaged over all RWC configurations, $P_d(r,t)$,
we note that it is related to $p(\ell,t)$ by 
\begin{equation}\label{Prtphi}
P_d(r,t)=\int_0^\infty d\ell\,\Phi_d(r,\ell)p(\ell,t)
\end{equation}
and is normalized according to $\int d^dr\,P_d(r,t)=1$. 
Another possibility is a normalization on the RWC fractal,
{\it i.e.,} $\int_0^\infty dr\,r^{d_f-1}Q_d(r,t)=1$. Both
distributions are simply related to each other by
$P_d(r,t)=\rho(r)Q_d(r,t)$.

In Eq.~(\ref{Prtphi}), $\Phi_d(r,\ell)$ represents the probability
for a site $r$ to belong to a RWC at distance $\ell$ from the origin
along the chain.  The chemical distance $\ell$ plays the role of the
time variable in Eq.~(\ref{Plt}), and one can immediately write
\begin{equation}\label{Phifin}
\Phi_d(r,\ell)=A_d\left({1\over2\pi\ell}\right)^{d/2}
   \exp\left(-{r^2\over2\ell}\right)
\end{equation}
where $A_d$ is a normalization factor such that
$\int d^dr\,\Phi_d(r,\ell)=1$.  Therefore, by inserting (\ref{Plt})
and (\ref{Phifin}) in (\ref{Prtphi}) we infer \cite{Fractal1,Note1}
\begin{equation}\label{Prt}
P_d(r,t)=\left({1\over2\pi}\right)^{d/2}{2 A_d\over(2\pi t)^{1/2}}
 \int_0^\infty d\ell\,\ell^{-d/2}\exp\left(-{r^2\over2\ell}\right)
\exp\left(-{\ell^2\over2t}\right).
\end{equation}
Now, the elementary transformation $x=\ell/r^2$ brings (\ref{Prt})
to the form
\begin{equation}\label{Prtsca}
P_d(r,t)=2 A_d (2\pi)^{-(d+1)/2}r^{-d}f_d(\xi),
\end{equation}
where the scaling function $f_d(\xi)$ is defined by
\begin{equation}\label{fdxi}
f_d(\xi)=\xi^2\int_0^\infty dx\,x^{-d/2}
  \exp\left[-{1\over2}\left(\xi^4x^2+{1\over x}\right)\right]
\end{equation}
for the scaling variable $\xi\equiv r/t^{1/4}$.  If the RWC
normalization is chosen, the distribution 
$Q_d(r,t)=\rho^{-1}(r)P_d(r,t)\cong t^{-1/2}{\tilde f}_d(\xi)$, where 
${\tilde f}_d(\xi)=\xi^{-2}f_d(\xi)$.

To deal now with the evaluation of $f_d(\xi)$ when $\xi\to0$, it is
convenient to rewrite the integrand exponent as
\begin{displaymath}
\exp\left[-{1\over2}\left(\xi^4x^2+{1\over x}\right)\right]=
  \exp\left[-{1\over2}\left(\xi^4x^2+{1\over x^2}\right)\right]
  \exp\left[-{1\over2x}\left(1-{1\over x}\right)\right]
\end{displaymath}
and expand the second exponential factor in Taylor series. The
remaining integrals can be solved exactly (see, {\it e.g.\/}
\cite{GrRy}), and one arrives at the following expression for
(\ref{fdxi})
\begin{equation}\label{Prtser}
f_d(\xi)=\xi^2\sum_{n=0}^\infty{1\over n!}\left(-{1\over2}\right)^n
  \sum_{k=0}^n (-1)^k{n\choose k}\xi^{d/2-1+n+k}
  K_{{1\over2}(d/2-1+n+k)}(\xi^2),
\end{equation}
where $K_\nu$ is the modified Bessel function of order $\nu$.

Let us consider Eq.~(\ref{Prtser}) in some particular cases
of interest.  The results for spatial dimensions $d\le7$ are
summarized in Table~\ref{table1}.  All the coefficients were
calculated numerically by computing the double sums explicitly.
In some cases they are available in analytic form, but we
include their numerical values to make the table uniform.

However, besides the coefficients, the main properties of these
expansions can be obtained readily as follows.  The key parameter
is $s={1\over2}({d\over2}-1)$.  The corresponding values of
$d=2(2s+1)$ for integer $s$ should be referred to as {\it critical
dimensions,\/} $d_c=2,6,10,\ldots$.  Each order in the expansion
has its own critical dimension.  The leading term has $d_c=2$, the
first correction term has $d_c=6$, the second correction term has
$d_c=10$, {\it etc.\/}  This has to do with the functional form of
$f_d(\xi)$ in the corresponding order which for $d<d_c$ depends on
$d$, at $d=d_c$ it has a logarithmic correction and for $d>d_c$
becomes independent of $d$.  In particular, the leading term of
$f_d(\xi)$ behaves as $\xi$ for $d=1$ and $\xi^2\ln(1/\xi)$ for
$d=d_c=2$ and as $\xi^2$ for all $d>d_c=2$, the first correction
term behaves as $\xi^d$ for $2\leq d<6$, and for $d=d_c=6$ as
$\xi^6\ln(1/\xi)$ and as $\xi^6$, for all $d>d_c=6$, and so on.

Mathematically, this behavior can be explained by the intrinsic
properties of the Bessel function $K_s(\xi^2)$.  By its
definition, $K_s(\xi^2)={\pi\over2}{\rm cosec}(\pi s)
[I_s(\xi^2)-I_{-s}(\xi^2)]$ for non-integer $s$ and, in turn,
$I_s(\xi^2)=\xi^{2s}\sum_{k=0}^\infty b_k(s)\xi^{4k}$ and 
$\xi^{2s}I_{-s}(\xi^2)=\sum_{k=0}^\infty b_k(-s)\xi^{4k}$.  As
one can see this expansion has {\it s-independent\/} powers
of $\xi$ that form the {\it invariant\/} part of $f_d(\xi)$.
The first terms of $\xi^{2s}K_s(\xi^2)$ expansion are
\begin{displaymath}
\xi^{2s}K_s(\xi^2)\approx b_0(s)\xi^{4s}+b_1(s)\xi^{4s+4}+\ldots
  -b_0(-s)-b_1(-s)\xi^4-\ldots .
\end{displaymath}
Thus, for $0<s<1$ ({\it i.e.,\/} $2<d<6$) $f_d(\xi)$ has the
form $f_d(\xi)\approx\xi^2[a_0(-d)-a_0(d)\xi^{d-2}]$.
We see that the first term of this expansion is the invariant
part (up to numerical coefficients) of $f_d(\xi)$, which
remains unchanged when varying $d$.  The same argument shows
that for $1<s<2$ ($6<d<10$) $f_d(\xi)$ takes the form
$f_d(\xi)\approx\xi^2[a_0(-d)-a_1(-d)\xi^4+a_0(d)\xi^{d-2}]$
and now {\it two\/} first terms of this expansion are the
invariant part of $f_d(\xi)$.  A special case in our problem
is $d=1$, {\it i.e.,\/} $s=-{1\over4}<0$.  Then the leading
term of $f_d(\xi)$ is $\xi^2\xi^{-4|s|}=\xi$, which is easily
seen by noting that $K_{-s}(\xi^2)=K_s(\xi^2)$.

For integer values of $s$, the small-$\xi$ expansion of
$\xi^{2s}K_s(\xi^2)$ has a logarithmic term of form
$\xi^{4s}\log(1/\xi)=\xi^{d-2}\log(1/\xi)$ \cite{GrRy}.

In general, there are $[s]+1$ ($[s]$ is the integer part of $s$)
terms in the invariant part of $f_d(\xi)$.

\section{Conclusions}

We have studied analytically the small-$\xi$ expansion of the
mean probability density, $P_d(r,t)$, of random walks on
random-walk chains in $d$-dimensional space.  We have shown
that the leading terms of the expansion of, $P_d(r,t)$, behaves,
in the limit $\xi=r/t^{1/4}\to0$, as
$$P_d(r,t)\propto\rho(r)t^{-1/2}(1-a_d\xi^{d-2}),
                      \qquad {\rm when} \quad 3\le d\le 5, $$
and as
$$P_d(r,t)\propto\rho(r)t^{-1/2}(1-c_d\xi^4),
                             \qquad {\rm when} \quad d\ge7,$$
where $\rho(r)\sim r^{d_f-d}$ and $d_f=2$. This implies that
the probability density $Q_d(r,t)=P_d(r,t)/\rho(r)$ on the fractal
chain behaves for $d\geq7$ as $Q_d(r,t)\sim t^{-1/2}(1-c_dr^4/t)$,
consistent with the behavior of diffusion in $\ell$-space,
{\it i.e.,\/} $p(\ell,t)\sim t^{-1/2}(1-\ell^2/2t)$, for
$\ell^2\ll t$, and the fact that $Q_d(r=1,t)\sim p(\ell=1,t)$ when
$d\to\infty$.  We see that this already occurs when $d\ge7$.

We have shown that logarithmic corrections occur at critical
dimensions $d=d_c=4 n+2$, with $n=0,1,2,\ldots$, {\it i.e.,\/}
$d_c=2,6,10,\ldots$, for the terms $\xi^{d_c-2}\log(1/\xi)$.
In particular for $d=2$, $Q_2(r,t)\simeq t^{-1/2}\ln(t^{1/4}/r)$,
for $r\ll t^{1/4}$, and the probability density for the random walker
to be close to the origin, $Q_2(r,t)$ behaves as $t^{-1/2}\ln t$.
This logarithmic correction is due to the fact that in two
dimensions the RWC returns to its starting point with probability
one.  In one dimension, $Q_1(r,t)\simeq t^{-1/4}/r$, and in $d=3$,
$Q_3(0,t)\simeq t^{-1/2}$.  One can say that $d=2$ plays the role
of a marginal dimensionality for the probability density of being
at the origin of random walks on RWC, while for larger $r$ each
order in the expansion has its own critical dimension.
\bigskip

{\bf Acknowledgments:}

We like to thank Julia Dr\"ager, Markus Porto and Grisha
Berkolaiko for valuable discussions.  Financial support by
the Alexander-von-Humboldt foundation, and the Deutsche
Forschungsgemeinshaft is gratefully acknowledged.

\widetext 
\begin{table}
\caption{The leading and correction terms for the series
expansion of $f_d(\xi)$, with $\xi\equiv r/t^{1/4}$, when $\xi\to0$,
Eq.~(\protect{\ref{Prtser}}), as a function of dimension $d$.}

\begin{tabular}{cccc}
  \multicolumn{1}{c}{$d     $}&
  \multicolumn{1}{c}{Leading term     }&
  \multicolumn{1}{c}{First correction }&
  \multicolumn{1}{c}{Second correction}\\
\tableline
 1 &$2.1558~\xi        $&$-2.5066~\xi^2     $&$O(\xi^5)       $\\
 2 &$2\xi^2\ln(1/\xi)  $&$0.1738\xi^2       $&$O(\xi^6)       $\\
 3 &$2.5066~\xi^2      $&$-0.0609~\xi^3     $&$O(\xi^6)       $\\
 4 &$2~\xi^2           $&$-1.2533~\xi^4     $&$O(\xi^6)       $\\
 5 &$2.5066~\xi^2      $&$-1.4372~\xi^5     $&$O(\xi^6)       $\\
 6 &$4~\xi^2           $&$-\xi^6\ln(1/\xi)  $&$-0.3369\xi^6   $\\
 7 &$7.5199~\xi^2      $&$-1.2533~\xi^6     $&$ 0.8244~\xi^7  $\\
\end{tabular}
\label{table1}
\end{table}
\narrowtext
 

\begin{thebibliography}{99}

\bibitem{Orbach} S.~Alexander and R.~Orbach, {\it J.~Phys. Lett.\/}
                 {\bf 43}, L625 (1982)

\bibitem{Havlin} S.~Havlin and D.~Ben-Avraham, {\it Adv. Phys.\/}
                 {\bf 36}, 695 (1987)

\bibitem{Kehr} K.~W.~Kehr and R.~Kutner, {\it Physica\/} A{\bf 110},
               535 (1982)

\bibitem{Fractal1} A.~Bunde and S.~Havlin, eds.,
                {\sl Fractals and disordered systems\/}
                (Springer Verlag, Heidelberg, 1991)

\bibitem{Guyer} R.~A.~Guyer, {\it Phys. Rev.\/} A{\bf 29}, 2751 (1984)

\bibitem{Procaccia} B.~O'Shaughnessy and I.~Procaccia, 
                    {\it Phys. Rev. Lett.\/} {\bf 54}, 455 (1985);
                    J.~P.~Bouchaud and A.~Georges, {\it Phys. Rep.\/}
                    {\bf 195} (1990)

\bibitem{BHR} A.~Bunde, S.~Havlin, and H.~E.~Roman, {\it Phys. Rev.\/}
              A{\bf 42}, 6274 (1990)
              
\bibitem{Blumen} J.~Klafter, G.~Zumofen and A.~Blumen, {\it J.~Phys.\/}
                 A{\bf 24}, 4835 (1991)

\bibitem{Roman92} H.~E.~Roman and M.~Giona, {\it J.~Phys.\/} A{\bf 25},
                  2107 (1992); M.~Giona and H.~E.~Roman, {\it Physica\/}
                  A{\bf 185}, 87 (1992)

\bibitem{Aharony92} A.~Aharony and A.~B.~Harris, {\it Physica\/} A{\bf 25},
                    (1992); G.~H.~Weiss, {\sl Aspects and Applications of
                    the Random Walk\/} (North Holland, Amsterdam, 1994)

\bibitem{Roman95} H.~E.~Roman, {\it Phys. Rev.\/} E{\bf 51}, 5422 (1995)

\bibitem{Yuste95} S.~B.~Yuste, {\it J. Phys.\/} A{\bf 28}, 7027 (1995)

\bibitem{DRB95} J.~Dr\"ager, S.~Russ and A.~Bunde,
                {\it Europhys. Lett.\/} {\bf 31}, 425 (1995)

\bibitem{Note1}
Actually, the function $\Phi_d(r,\ell)=0$ when $\ell<\ell_{\rm min}$,
and $\ell_{\rm min}=r$ when all RWC configurations are considered (see
\cite{Julia}).  Thus, the lower integration limit in Eq.~(\ref{Prt})
corresponds to $\ell_{\rm min}=r$.  However, by performing the
transformation $x=\ell/t^{1/2}$, it can be shown to vanish, for fixed
$\xi=r/t^{1/4}$, as $\xi/t^{1/4}\to0$ when $t\to\infty$.  Hence, we set
the lower integration limit $\ell_{\rm min}=0$ in Eq.~(\ref{Prt}).

\bibitem{GrRy} I.~S.~Gradshteyn and I.~M.~Ryzhik, {\sl Table of Integrals,
               Series and Products\/} (Academic Press, New York 1965)

\bibitem{Julia} A.~Bunde and J.~Dr\"ager, {\it Phys. Rev.\/} E{\bf 52}, 53 
                (1995)

\end{thebibliography}
\end{document}